\documentclass[12pt,a4paper,tightenlines,nofootinbib]{revtex4}

\usepackage{amssymb}
\usepackage{amsmath}
\usepackage{epsf}
\usepackage{psfrag}

\def\nablaslash{\not{\hbox{\kern-3pt $\nabla$}}}

\begin{document}

\author{Jaume Garriga$^{1,2}$ and Ariel Megevand$^{1,2}$}
\affiliation{$^1$ Departament de F{\'\i}sica Fonamental,
Universitat de Barcelona, Diagonal 647, 08028 Barcelona, Spain}
\affiliation{$^2$ IFAE, Campus UAB, 08193 Bellaterra (Barcelona),
Spain}
\title{Decay of de Sitter vacua by thermal activation}
\date{\today}

\begin{abstract}

Decay of a de Sitter vacuum may proceed through a ``static"
instanton, representing pair creation of critical bubbles
separated by a distance comparable to the Hubble radius -- a
process somewhat analogous to thermal activation in flat space. We
compare this with related processes recently discussed in the literature.

\end{abstract}

\maketitle

\section{introduction}

In field theory, a metastable false vacuum may decay either by
tunneling or by thermal activation. Tunneling is described by a
solution of the Euclidean field equations symmetric under
spacetime rotations \cite{coleman}. In flat space
and at zero temperature this is the instanton with the least
action, and hence represents the dominant contribution to the
decay rate \cite{theorem}. At sufficiently high temperature,
thermal activation is more probable than tunneling, and the rate
is dominated by a static solution which represents a spherical
critical bubble in unstable equilibrium between expansion and
collapse \cite{fintemp}. The symmetry of this solution is
O(3)$\times$ U(1), where the U(1) factor corresponds to
translations along the compactified Euclidean time direction.

As shown by Coleman and de Luccia \cite{deLuccia}, gravity can be
easily incorporated into the description of tunneling. When the
initial state is a false vacuum with a positive energy density,
the initial geometry corresponds to a de Sitter-like exponential
expansion. A bubble which materializes through quantum tunneling
has zero energy, and consequently (in the thin wall limit) the
geometry outside of the bubble remains unaffected by the
nucleation event. After nucleation, the bubble walls accelerate
outward, and the volume of the new phase increases at the expense
of the old one. However, due to the presence of event horizons in
de Sitter, the growth of a single bubble cannot engulf the whole
space. If the nucleation rate per unit volume $\Gamma$ is large
compared to $H^4$, where $H$ is the inverse de Sitter radius, the
foam of nucleated bubbles will percolate and the phase transition
will complete. But if $\Gamma\ll H^4$, the rate at which bubbles
nucleate and grow does not catch up with the exponential expansion
of the false vacuum. In this case, the volume of the false vacuum keeps
increasing with time and the transition never fully completes in the whole spacetime (leading to eternal old inflation). However, any observer will experience a local transition to the new vacuum phase in a finite proper time.

On the other hand, the description of thermal activation can be more involved when
the self-gravity of the bubbles and of the thermal bath is
considered. In the cosmological context, the thermal bath drives
the expansion of the universe, and the temperature becomes time
dependent. Because of that, exact instantons cannot be
constructed. In low energy cosmological phase transitions (e.g. at
the electroweak scale) it is safe to ignore the self-gravity of
the bubbles, and to use the flat spacetime results for the
nucleation rate\footnote{The effects of self-gravity, however,
may be important at high energy scales, see e.g. \cite{Donoghue}
for a recent discussion.}. If this rate is sufficiently
large, the phase transition completes as the bubbles percolate,
typically after a short period of supercooling and subsequent
release of latent heat by the nucleated bubbles \cite{ariel}. But
if the rate is too small, there is a strong
supercooling and the thermal bath dilutes away; the
false vacuum starts dominating before the
phase transition is complete, and we are back to the situation
described at the end of the previous paragraph \cite{guwe}.

Note, however, that even when all matter has been diluted away, the false
vacuum dominated de Sitter expansion can be considered to have a
non-vanishing temperature \cite{giha} $T=H/2\pi$, which does not
dilute further as the universe expands. One may then ask whether
this leads to thermal activation, and if so, at what rate does it
proceed. In what follows, we shall discuss the corresponding
instanton, describing the nucleation of a pair of critical bubbles
in unstable equilibrium between expansion and collapse. It should
be noted that the theorem in Ref. \cite{theorem} does not
necessarily apply to de Sitter space, and hence it is not clear a
priori whether this new instanton has higher action than the usual
Coleman-de Luccia one.

Static self-gravitating instantons with O(3) symmetry have
previously been considered in a variety of contexts, notably for
the description of false vacuum decay in the presence of a black
hole (See e.g. \cite{baby,hiscock,bkk} and references therein). The
particular solution we shall consider here corresponds to pair
creation of critical bubbles in de Sitter, and to our knowledge it
does not seem to have received much attention in the past. The
following is an extended version of the discussion given by the
present authors in \cite{game}. In Section II we describe the
solution. In Section III we discuss the action and the nucleation
rate. In Section IV we consider the limit in which the mass of the
nucleated bubbles is small. Section V is devoted to the opposite
limit, when the gravity of the bubbles is very important. In
Section VI we compare the action for thermal activation with the
action for tunneling (through the Coleman-de Luccia instanton). Section VII compares the process of thermal activation of seeds of the new phase with a related
process recently discussed by Gomberoff et al. \cite{gomberoff}, by which most of space would jump to the new phase except for a pair of bubbles which contain the ``remnant" of the old phase. Section VIII is devoted to conclusions.

\section{Pair creation of critical bubbles}

\begin{figure}[tbh]
\psfrag{rm}[][r]{$r_+$}\psfrag{R}[][r]{$R$}\psfrag{0}[][r]{$0$}
\psfrag{t}[][r]{$t$} \psfrag{f}[][l]{$\phi$} \centering
\epsfysize=5cm \leavevmode \epsfbox{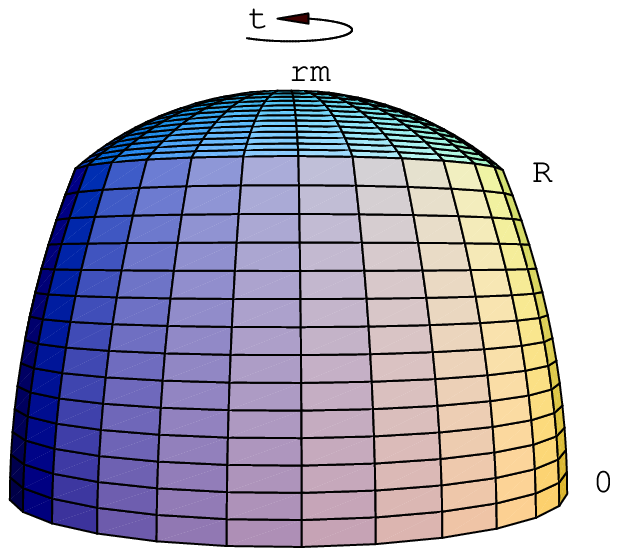}
\epsfbox{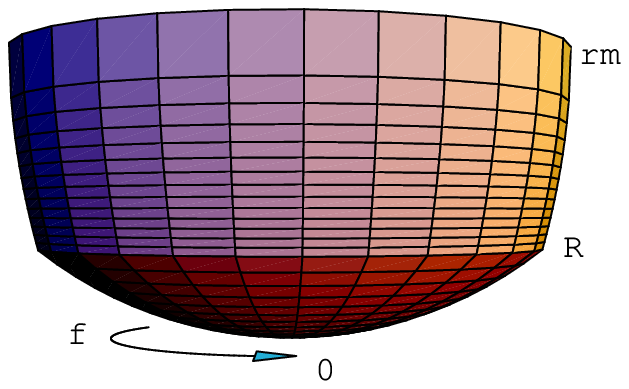} \caption{Static instanton in de
Sitter space. The left figure shows the geometry induced on the
plane $r,t$, while keeping angular coordinates fixed, whereas the
right figure shows the geometry induced on the plane $r,\phi$,
keeping $\theta$ and $t$ fixed. The vertical direction corresponds
to the coordinate $r$, common to both pictures. The cosmological
horizon is at $r=r_+$, the buble wall is at $r=R$, and $r=0$ is
the center of the static bubble of the new phase. The geometry at
the time of nucleation is obtained by cutting the instanton by a
smooth spacelike surface orthogonal to the time-like killing
vector. This corresponds to a diametral section of the figure on
the left, which therefore contains two bubbles, whose centers are
separated by a distance comparable to the Hubble radius. }
\label{estatico}
\end{figure}

Unlike the case of the Coleman-de Luccia bubble, the energy of a
critical bubble is different from zero, and consequently, the
metric outside of it is no longer pure de Sitter but
Schwarzschild-de Sitter (SdS). The instanton is a solution of the
Euclidean equations of motion, with two metrics glued together at
the locus of the wall, which is a surface of constant $r$ in the
static chart of SdS (see Fig.~\ref{estatico}). For simplicity, we
shall restrict attention to the case where the vacuum energy
density is positive in both the initial and the final states. Also, we shall assume that the thin wall approximation is valid \cite{coleman}.

\subsection{The instanton}

The metric outside is given by
\begin{equation}
ds^{2}=f_{o}\left( r\right) dt^{2}+f_{o}^{-1}\left( r\right)
dr^{2}+r^{2}d\Omega ^{2},  \label{metrico}
\end{equation}
where $d\Omega ^{2}=d\theta ^{2}+\sin ^{2}\theta d\phi ^{2}$, and
\begin{equation}
f_{o}(r)=\left( 1-\frac{2GM}{r}-H_{o}^{2}r^{2}\right) .
\label{efeo}
\end{equation}
The metric inside is given by
\begin{equation}
ds^{2}=C^{2}f_{i}\left( r\right) dt^{2}+f_{i}^{-1}\left( r\right)
dr^{2}+r^{2}d\Omega ^{2},  \label{metrici}
\end{equation}
where
\begin{equation}
f_{i}\left( r\right) =\left( 1-H_{i}^{2}r^{2}\right).
\label{efei}
\end{equation}
Here $G$ is Newton's
constant. The parameter $C$ is determined by the condition that on
the bubble wall (i.e., at $r=R$) the two metrics must agree, which
leads to $C=[f_{o}\left( R\right) /f_{i}\left( R\right)]^{1/2} $.

The parameters $M$ and $R$ depend on the wall tension $\sigma$,
and the Hubble parameters outside and inside the bubble, $H_{o}$
and $H_{i}$ respectively. Their values are determined by the
junction conditions at the bubble wall \cite{israel},
\begin{equation}
\left[ K_{ab}\right] =-4\pi G\sigma \gamma _{ab},
\label{discont2}
\end{equation}
where $\left[ K_{ab}\right] $ is the difference in the extrinsic
curvature $K_{ab}=(1/2)f^{1/2}\partial _{r}g_{ab}$ on the two
sides and $\gamma _{ab}$ is the world-sheet metric. Eq.\
(\ref{discont2}) gives rise to the junction conditions,
%\begin{eqnarray}
%\left[ \frac{1}{2}f^{-1/2}f^{\prime }\right] &=&-4\pi G\sigma , \\
%\frac{1}{r}\left[ f^{1/2}\right] &=&-4\pi G\sigma .
%\end{eqnarray}
\begin{equation}
[g]=-4\pi G \sigma,\quad [g']=0, \label{juco}
\end{equation}
where we have introduced the new function $g(r)=f^{1/2}(r)/r$.
Using Eqs.~\ (\ref{efeo}) and (\ref{efei}), we have
\begin{equation}
g_{o}g_o^{\prime } =-\frac{1}{r^{3}}+\frac{3GM}{r^{4}}, \quad\quad
g_{i}g_i^{\prime } =-\frac{1}{r^{3}}. \label{cuadradosprima}
\end{equation}
From (\ref{juco}), we have $g_{o}^{\prime }\left( R\right)
=g_{i}^{\prime }\left( R\right)= -3M/4\pi \sigma R^{4}$, and then
$g_{i}(R)$ and $g_{o}(R)$ are easily obtained from
Eqs.~(\ref{cuadradosprima}):
\begin{equation}
g_{i}\left( R\right) =\frac{4\pi \sigma R}{3M}, \quad\quad
g_{o}\left( R\right) =g_{i}\left( R\right) \left(
1-\frac{3GM}{R}\right) . \label{ges}
\end{equation}
From (\ref{efeo}) and (\ref{efei}) we have
\begin{equation}
g_o^{2 }(R) =\frac{1}{R^{2}}-\frac{2GM}{R^{3}}-H_{o}^{2},
\quad\quad g_i^{2 }(R) =\frac{1}{R^{2}}-H_{i}^{2}.
\label{cuadrados}
\end{equation}
Inserting (\ref{ges}) in (\ref{cuadrados}) we finally obtain a
quadratic equation for $g_{i}\left( R\right) \equiv x$. The
solution is
\begin{equation}
x=\frac{\epsilon }{4\sigma }+\frac{3\sigma }{16M_{p}^{2}}+\left[
\left( \frac{\epsilon }{4\sigma }+\frac{3\sigma
}{16M_{p}^{2}}\right) ^{2}+\frac{ H_{i}^{2}}{2}\right] ^{1/2},
\label{x}
\end{equation}
where we have introduced the parameter $\epsilon$ representing the
difference in vacuum energy difference on both sides of the bubble
wall: $H_{o}^{2}-H_{i}^{2}=8\pi G\epsilon /3=\epsilon
/3M_{p}^{2}$. Then the parameters $M$ and $R$ are given in terms
of $x$ by
\begin{equation}
R^{-2} =x^{2}+H_{i}^{2}, \quad\quad M =4\pi \sigma R/3x.
\label{themass}
\end{equation}
This concludes the construction of the instanton solution for
given values of the physical parameters $\sigma, H_o$ and $H_i$.

\subsection{Cosmological horizon}

The above equations are valid only as long as $3GM<R$ [otherwise
(\ref{ges}) would yield $g_i<0$, which is meaningless]. Thus, from
(\ref{themass}), we require $4\pi G\sigma < x$, or
$\sigma<\sigma_{N}$, where
\begin{equation}
\sigma_{N}^2=4M_p^4 (3H_o^2-H_i^2). \label{sigmamax}
\end{equation}
(The case with $\sigma>\sigma_{N}$ will be discussed in Section V.)
The mass parameter satisfies
\begin{equation}
M \leq M_{N} \equiv M(\sigma_{N}) = (3\sqrt{3} GH_o)^{-1}.
\label{mmax}
\end{equation}
For $M<M_{N}$, the equation $f_{o}\left( r\right) =0$ has
three real solutions. One of them, say $r_-$, is negative and the
other two are positive. The two positive roots correspond to the
black hole and cosmological horizons. We call them respectively
$r_s$ and $r_{+}$. Therefore we can write
\begin{equation}
f_{o}\left( r\right) =-\frac{H_{o}^{2}}{r}\left( r-r_-\right)
\left( r-r_s\right) \left( r-r_{+}\right) . \label{fofactors}
\end{equation}
In the present case, with $\sigma<\sigma_{N}$, the horizon at $r_s$ is not present,
since the exterior metric is matched to an interior metric at some
$r=R > r_s$ (see Fig. \ref{estatico}). For $r<R$ the metric is
just a ball of de Sitter in the static chart, and it is regular
down to the center of symmetry at $r=0$. In general, the size of
the cosmological horizon is given by
%\begin{equation}
%H_{o}r_{+}=\frac{3^{1/3}+\left(
%-9H_{o}MG+\sqrt{-3+81H_{o}^{2}M^{2}G^{2}} \right)
%^{2/3}}{3^{2/3}\left( -9H_{o}MG+\sqrt{-3+81H_{o}^{2}M^{2}G^{2}}
%\right) ^{1/3}}.
%\end{equation}
%This expression is rather cumbersome for analyzing the limiting
%cases,
%[and even to see that it gives a real answer for $3GM<\left(
%\sqrt{3}H_{o}\right) ^{-1}$, since the square root is purely
%imaginary].
%and it is convenient to introduce the angle
\begin{equation}
r_{+}=\frac{2
H_o^{-1}}{\sqrt{3}}\cos\left(\frac{\varphi+\pi}{3}\right),
\end{equation}
where we have introduced the angle
\begin{equation}
\varphi=- \arctan \sqrt{{1\over 27 H_o^2 M^2 G^2}-1},
\label{varphi}
\end{equation}
In the limit $M\rightarrow 0$ the angle
$\varphi\rightarrow-\pi/2$, and $H_{o}r_{+}\rightarrow 1$.

According to Eq\ (\ref{ges}), on the bubble wall we have $f_{o}\left(
R\right) =x^{2}\left( R-3GM\right) ^{2}$, so the equation
$f_{o}\left( R\right) =0$ has a double zero instead of two
different roots. This means that the radius of the instanton will
coincide with the radius of one of the horizons only in the
special case where both horizons have the same size, $
r_s=r_{+}=R=3GM$. As we shall see in Section V,
this limit corresponds to $\sigma=\sigma_{N}$, for which the exterior metric
is the Nariai solution \cite{nariai,perry}, with mass
parameter $M_{N}$ given in (\ref{mmax}) and with
$H_{o}r_{+}\rightarrow 1/\sqrt{3}$. There, we shall
also comment on the case $\sigma>\sigma_{N}$, which is not
covered by the present discussion.

\subsection{Euclidean periodicity}

Regularity of the Euclidean solution will determine the
periodicity of the Euclidean time coordinate (and the thermal properties of the solution). For $r\rightarrow r_{+}$,
we have
\begin{equation}
f_o(r)\approx A^{2}\left( 1-\frac{r}{r_{+}}\right),
\end{equation}
where
\begin{equation}
A^{2}=H_{o}^{2}\left( r_{+}-r_-\right) \left( r_{+}-r_s\right)
=3H_{o}^{2}r_{+}^{2}-1 . \label{A}
\end{equation}
In terms of the new coordinates
\begin{equation}
\rho =\frac{2r_{+}}{A}\sqrt{1-\frac{r}{r_{+}}},\quad \phi
=\frac{A^{2}}{ 2r_{+}}t,
\end{equation}
the metric (\ref{metrico}) for $r\to r_+$ reads
\begin{equation}
ds^{2}=\rho ^{2}d\phi ^{2}+d\rho ^{2}+r_{+}^{2}d\Omega ^{2},
\end{equation}
so it is clear that $\phi $ is an angle, $0\leq \phi \leq 2\pi$,
and $t$ varies in the range $0\leq t\leq 4\pi r_{+}/A^{2}$.
Thus, the periodicity of the time coordinate is given by
\begin{equation}
\beta =\frac{4\pi r_+}{3 H_{o}^2 r_{+}^{2}-1}= \frac{2\pi
r_+^2}{r_{+}-3GM}.
\end{equation}
It is also of some interest to determine the physical temperature
on the bubble wall worldsheet, given by the proper time
periodicity $ \beta _{R}\equiv \int_{0}^{\beta }f_{o}^{1/2}\left(
R\right) dt=f_{o}^{1/2}\left( R\right) \beta =Cf_{i}^{1/2}\left(
R\right) \beta $,
\begin{equation}
\beta _{R}=2\pi xr_{+}^{2}\frac{R-3GM}{r_{+}-3GM}.  \label{betar}
\end{equation}
If there are field degrees of freedom living on this worldsheet,
this will be the temperature that they will experience (rather
than the ambient de Sitter temperatures).

Like in the case of instantons describing the production of black
holes \cite{perry} or monopoles \cite{basu} in de Sitter, the
instanton presented here describes the creation of {\em pairs} of
bubbles. As we have just seen, the Euclidean time runs on a circle $S^1$ (See
Fig. 1). The geometry at the time of nucleation is obtained by
slicing the compact instanton through a smooth spacelike surface
which cuts the $S^1$ factor at two places, say, $t=0$ and
$t=\beta/2$. The resulting geometry contains two different bubbles
separated by a distance comparable to the inverse expansion rate.

\section{Instanton action}

The nucleation rate is determined by the Euclidean action, which
turns out to have a rather simple expression in terms of $r_+$.
The action is given by
\begin{equation}
S_{E}=\sigma \int d^{3}\xi \sqrt{\gamma }+\int
d^{4}x\sqrt{g}\left( \rho _{V}-\frac{\mathcal{R}}{16\pi G}\right).
\end{equation}
By the equations of motion, the scalar curvature is given by
\begin{equation}
\mathcal{R}\sqrt{g}=32\pi G\rho _{V}\sqrt{g}+24\pi G\sigma \int
d^{3}\xi \sqrt{\gamma }\delta ^{(4)}\left( x-x\left( \xi \right)
\right),
\end{equation}
and hence the on shell action is given by
\begin{equation}
S_{E}=-\frac{\sigma }{2}\int d^{3}\xi \sqrt{\gamma }-\int
d^{4}x\rho _{V}\sqrt{g}.  \label{actionstatic}
\end{equation}
The first integral in (\ref{actionstatic}) is just the volume of a
two-sphere of radius $R$ times $\beta_R$. The second integral in
(\ref{actionstatic}) splits into the contributions from the two different vacua,
\begin{eqnarray}
&&\rho _{i}\int_{0}^{R}Cdtdr4\pi r^{2}+\rho
_{o}\int_{R}^{r_{+}}dtdr4\pi
r^{2} \\
&=&\rho _{i}C\beta \frac{4}{3}\pi R^{3}+\rho _{o}\beta
\frac{4}{3}\pi \left( r_{+}^{3}-R^{3}\right)
\end{eqnarray}
So the instanton action is
\begin{equation}
S_{E}=-2\pi R^{2}\sigma f_{o}^{1/2}\left( R\right) \beta
-R^{3}\frac{H_{i}^{2}}{ 2G}\frac{f_{o}^{1/2}\left( R\right)
}{f_{i}^{1/2}\left( R\right) }\beta -\left( r_{+}^{3}-R^{3}\right)
\frac{H_{o}^{2}}{2G}\beta .
\end{equation}
After some algebra $S_{E}$ can be written in the simple form
\begin{equation}
S_{E}=-\frac{\pi r_{+}^{2}}{G} = -{A(r_+)\over 4 G}, \label{area}
\end{equation}
where $A(r_+)$ is the area of the horizon at $r_+$. The exponent
$B$ which gives the probability for brane nucleation is  the
difference in actions between instanton and background.
The action of the background is just $ S_{E}= -\pi/ GH_{o}^{2}, $
so the difference in actions between the instanton and
the background is given by
\begin{equation}
B=\frac{\pi }{GH_{o}^{2}}\left( 1-r_{+}^{2}H_{o}^{2}\right).
\label{B}
\end{equation}
This expression could have been anticipated from general
arguments. The action of a static Euclidean solution with
periodicity $\beta$ is given by \cite{bateza,hoha} $S_E= \beta E -
S$, where $E$ is the {\em total} energy and $S$ is the entropy.
For a closed solution, without a boundary, the total energy
vanishes $E=0$. Hence, the Euclidean action is equal to the
entropy of the solution, which is just one fourth of the sum of
the areas of all horizons. Hence, Eq. (\ref{B}) can be rewritten
as $B=-\Delta S$, and the transition probability is proportional
to
\begin{equation}
\Gamma \sim \exp(\Delta S). \label{rateent}
\end{equation}
As both the initial and the final solutions can be said to
represent macroscopic states in thermal equilibrium, their entropy
can be interpreted as the logarithm of the corresponding number of
microstates in the microcanonical ensemble with $E=0$. The
transition probability is then simply proportional to the relative
number of microstates.

\section{Weak self-gravity limit}

If the bubbles are sufficiently light, $GMH_{o}\ll 1$, and provided that $\sigma<\sigma_N$, Eq.
(\ref{varphi}) gives $H_o r_+ \approx 1-GMH_{o}$, and from
(\ref{B}) we have
\begin{equation}
B\approx \beta_o M, \label{bm}
\end{equation}
where $\beta_o\equiv 2\pi/H_o$. Since the mass of the bubble is small, its appearance does not significantly change the
temperature of the horizon. In such case, the nucleation rate of
the bubbles may be interpreted from the point of view of the
observer at $r=0$ as being due to a thermal bath at the fixed
Gibbons-Hawking temperature $\beta_0^{-1}$. The corresponding
probability is proportional to the Boltzman factor
\begin{equation}
e^{-\beta_0 M}. \label{boltsu}
\end{equation}
Eq. (\ref{bm}) can also be understood as follows.
The energy of the bubble has been extracted from thermal
reservoir. According to the first law, the entropy of the
reservoir must decrease by $dS_{horizon} = -\beta_0\ dM$
\cite{claudio}. From the discussion at the end of the last
Section, $B=-(\Delta S)_{horizon}$, and hence for bubbles of small
mass we obtain (\ref{bm}).

Let us now consider a few specific limiting cases, starting with
the case of low tension branes, $\sigma/M_p^2  \ll H_{o},
H_o-H_i$, with $H_o> H_i$. In this case the parameter $x$ is large
compared with $H_{o}$, $R\simeq x^{-1}$, and we have
$$
M=16 \pi \sigma^3/3 \epsilon^2.
$$
This is just the flat space expression for the energy of a
critical bubble. The corresponding bounce action is $B\approx 32
\pi^2 \sigma^3/3 H_o \epsilon^2,$ which coincides with the thermal
activation rate in flat space at the temperature $\beta_o^{-1}$.

At finite temperature, jumps to a vacuum with a higher energy
density are also possible. In the absence of gravity, these jumps
are frustrated because the bubbles of the new phase tend to
recollapse. When gravity is included, the expansion of the
universe can keep these ``false vacuum" bubbles from contracting
(this is true also for the case of tunneling bubbles \cite{leewe}).
Hence, let us consider again the case of low tension branes,
$\sigma/M_p^2  \ll |H_o-H_i| \ll H_i$, but now with $H_o < H_i$.
In this case we find $x\approx -\sigma H_i^2/\epsilon \ll H_i$,
and
$$
R\approx H_i^{-1}\left[1-{1\over 2}\left({H_i\sigma\over
\epsilon}\right)^2\right]\approx H_i^{-1}.$$ In this limit, the
bubble of the false vacuum phase is almost as big as the
cosmological horizon. We also have $M\approx -(4\pi/3)\epsilon
H_i^{-3}$ and $B\approx -8\pi^2 \epsilon/3 H_o^{4}$, where we have
used that $H_i-H_o \ll H_i$ to replace $H_i$ by $H_o$ in the last
expression (note that $\epsilon<0$ in the case we are considering
here, so $M$ and $B$ are both positive). This approximately
coincides with the bounce action for the homogeneous Hawking-Moss
instanton \cite{hamo}, representing the upward jump of a horizon
sized region of de Sitter space into a higher false vacuum.

Finally, we may consider the case of intermediate tension
$|H_o-H_i| \ll \sigma/M_p^2 \ll H_o, H_i$. This leads to $x\approx
H_i/\sqrt{2}$, and $R^2 \approx 2/ 3H_i^2$. In this case, the
difference in pressure between inside and outside of the brane is
insignificant compared with the brane tension term, which is
balanced against collapse by the cosmological expansion. The
energy of the critical bubble is $E_c(R)\approx 4\pi \sigma R^2$.
Note, from (\ref{betar}), that the inverse temperature
\begin{equation}
\beta_R\approx {2\pi \over \sqrt{3}H_o},
\end{equation}
is different from the one experienced by a geodesic observer at
the origin of coordinates $r=0$. This is because observers at
$r\neq 0$ are in fact accelerating. From the point of view of the
observer at $r=0$, the energy of the bubble is $M= f_{0}^{1/2}(R)
E_c(R)$, because of the gravitational potential contribution.
Hence, taking into account that $\beta_R = f_{0}^{1/2}(R) \beta_0$
the exponent in the Boltzmann suppression factor can be written as
$B\approx\beta_0 M\approx \beta_R E_c$, and we have $B \approx 16
\pi^2 \sigma / 3 \sqrt{3}H^3_o$.

\section{Strong gravity limit}

For given $H_o$ and $H_i$, the solution of Section II only exists
provided that the tension of the bubble wall does not exceed a
certain bound $\sigma _{N}$, given in Eq. (\ref{sigmamax}).
Let us now consider what happens near this bound, and beyond.

\subsection{The Nariai limit}

As we mentioned in the discussion below Eq.~(\ref{fofactors}), the
exterior metric in the limit $\sigma\to \sigma_{N}$ corresponds
to the Nariai solution, with $r_s=r_{+}=\left(
\sqrt{3}H_{o}\right) ^{-1}$, and $M=1/3\sqrt{3} H_o G$. Replacing
this value in (\ref{B}) we find readily
\begin{equation}
B=\frac{2\pi }{3GH_{o}^{2}}.
\end{equation}
This may be compared with the action of the instanton describing
the nucleation of black holes in the same de Sitter universe \cite{perry},
\begin{equation}
B_N = \frac{\pi }{3GH_{o}^{2}}.
\end{equation}
The difference $B-B_N = \pi/3GH_o^2 = A_{bh}/4G$, is just the area
of the black hole horizon in the Nariai solution, as expected from
the general discussion of the previous Section (Here, we are of
course neglecting the entropy stored in the field degrees of
freedom living on the bubble walls, which would show up when the
determinantal prefactor in the nucleation rate is evaluated).

The fact that $r_s=r_{+}$
does not mean that both horizons coincide, since the coordinates
$r,t$ become inadequate in this case \cite{perry}. Near the Nariai limit
the metric outside takes the form (\ref{metrico}), with
\begin{equation}
f_o(r)\approx A^{2}\left( 1-\frac{r}{r_{+}}\right) -\left(
1-\frac{r}{r_{+}} \right) ^{2},
%dt^{2}+\frac{dr^{2}}{\left[
%A^{2}\left( 1-\frac{r}{r_{+}} \right) -\left(
%1-\frac{r}{r_{+}}\right) ^{2}\right] }+r_+^{2}d\Omega ^{2}
\end{equation}
and $r\approx r_+$, plus higher orders in the parameter $A$, which we defined in (\ref{A}). In the present limit this parameter tends
to zero, $A^{2}=\sqrt{3}H_{o}\left( r_{+}-r_s\right) $. Now we
define new coordinates $\psi$ and $\lambda$ by
\begin{equation}
\cos \psi =1-\frac{2}{A^{2}}\left( 1-\frac{r}{r_{+}}\right) ,\quad
\lambda =\frac{A^{2}}{2}t,
\end{equation}
so that the metric becomes
\begin{equation}
ds^{2}=\sin ^{2}\psi\ d\lambda ^{2}+r_{+}^2 d\psi
^{2}+r_{+}^2d\Omega ^{2}.  \label{nariai}
\end{equation}
The cosmological horizon is at $\psi =0$ and
the black hole horizon is at $\psi =\pi $. Now in the limit
$A\rightarrow 0$ we just replace $r_{+}=\left(
\sqrt{3}H_{o}\right) ^{-1}$.

We must determine the position $\psi _{R}$ of the bubble wall, which is
given as before by the matching conditions (\ref{discont2}), where
now the metric outside is (\ref{nariai}). So, on the wall, we
have
\begin{eqnarray}
ds_{\sigma }^{2} &=&\sin ^{2}\psi _{R}\ d\lambda
^{2}+r_{+}^{2}d\Omega ^{2} \\
&=&f_{i}\left( R\right) dt^{\prime 2}+R^{2}d\Omega ^{2}.
\end{eqnarray}
The extrinsic curvature on the outside is
$-(1/2)\partial _{\psi }g_{ab}$, with $g_{00}=\sin ^{2}\psi $ and
$ g_{\Omega \Omega }=r_{+}^{2}$, i.e., $ K_{00}
=-(1/r_{+})g_{00}\cot \psi , K_{\Omega \Omega } =0.$ The curvature
inside is as before $K_{00}=g_{00}
\partial_r f_{i}^{1/2}$ and $K_{\Omega \Omega }=g_{\Omega
\Omega }f_{i}^{1/2}/r$, with $ f_{i}\left( r\right) =\left(
1-H_{i}^{2}r^{2}\right) $, so the Israel conditions give
\begin{eqnarray}
-\frac{1}{r_{+}}\cot\psi_R -\left(
f_{i}^{1/2}\right) ^{\prime }|_R &=&-4\pi G\sigma , \\
f_{i}^{1/2}\left( R\right) /R &=&4\pi G\sigma .
\end{eqnarray}
These equations are easily solved and give
\begin{eqnarray}
\sin \psi _{R}  &=&\left( \frac{
3H_{o}^{2}-H_{i}^{2}}{6H_{o}^{2}-H_{i}^{2}}\right) ^{1/2}, \\
\sigma  = \sigma_{N} &=& 2M_{p}^{2}\sqrt{3H_{o}^{2}-H_{i}^{2}}
\label{sigmamax2}
\end{eqnarray}
so $H_{i}$ must be less than $\sqrt{3}H_{o}$. Now regularity at
the cosmological horizon $\psi \simeq 0$ implies that $0\leq
\lambda /r_{+}\leq 2\pi $, so $\beta _{R}=\sin \left( \psi
_{R}\right) 2\pi r_{+}$. Hence,
\begin{equation}
\beta _{R}=\frac{2\pi }{\sqrt{3}H_{o}}\left(
\frac{3H_{o}^{2}-H_{i}^{2}}{6H_{o}^{2}-H_{i}^{2}}\right) ^{1/2}.
\label{hite}
\end{equation}
Thus, also in this case, the effective temperature of the field
degrees of freedom living on the worldsheet will be of order $H_0$
(The only exception occurs if there is some fine adjustment
between $H_o$ and $H_i$ which makes the factor inside the brackets
very small, in which case the temperature may be much larger.)

\subsection{Beyond the Nariai limit}

For $\sigma > \sigma_{N}$ we have $3GM>R$ and the construction
of Section II does not apply [since by Eq. (\ref{ges}), $g_i$
would be negative]. As pointed out in \cite{gomberoff}, above this
threshold a static solution can still be constructed by gluing the
interior solution (\ref{metrici}) to the $r_s<r<R$ portion of the
exterior SdS solution (\ref{metrico}) (rather than using the
$R<r<r_+$ portion). This changes the sign of $K_{ab}$ in the
exterior, and the junction conditions become
\begin{equation}
\{g\}=4\pi G\sigma,\quad\quad \{g'\}=0, \label{average}
\end{equation}
where the curly brackets denote twice the average value on both
sides of the bubble wall. Eq. (\ref{ges}) is then replaced by
\begin{equation}
g_{i}\left( R\right) =\frac{4\pi \sigma R}{3M}, \quad\quad
g_{o}\left( R\right) =g_{i}\left( R\right) \left(
\frac{3GM}{R}-1\right) , \label{ges2}
\end{equation}
but Eqs. (\ref{x}) through (\ref{themass}) remain the same. The
instanton would still look pretty much as in Fig. 1, but with the
cosmological Horizon of radius $r_+$ replaced by a black hole
horizon of radius $r_s< R$. Hence, in the right pannel of Fig. 1,
the horizontal maximal circles would grow from $0$ to $R$ as we
move up from the center of the bubble, but then the circles would
start decreasing from $R$ to $r_s$ as we continue from the bubble
wall to the horizon.

It is straightforward to calculate the Euclidean action for this
solution, which is given by
\begin{equation}
S_E=-{A(r_s)\over 4 G},
\end{equation}
where $A(r_s)=4\pi r_s^2$ is the area of the black hole horizon,
with
$$
r_s= {2 H_o^{-1}\over\sqrt{3}}cos\left({\varphi-\pi \over
3}\right),
$$
and where $\varphi$ is given by (\ref{varphi}). The corresponding
bounce action
\begin{equation}
B=\frac{\pi }{GH_{o}^{2}}\left( 1-r_{s}^{2}H_{o}^{2}\right),
\label{B2}
\end{equation}
is perfectly finite, since the instantons involved are both
compact and regular. Moreover, $B>0$, as it should be if this is
to be interpreted as a process with an exponentially suppressed
rate.

Is this instanton suitable for describing vacuum decay in the
usual sense? Let us assume that we are in a false vacuum phase,
and for simplicity, that the false vacuum decay rate per unit
volume is exceedingly small compared with $H^4$. Then we expect
that after some time the metric will take the form
\begin{equation}
ds^2=-d\tilde t^2 + e^{2H_o \tilde t} (d\vec x)^2, \label{dsflat}
\end{equation}
over an exponentially large portion of space (with the exception
of small portions of volume carved out by bubbles of the new phase
which may have nucleated). In the solution described in Section
II, an asymptotic region with metric (\ref{dsflat}) can be found
(upon analytic continuation) in the region beyond the cosmological
horizon, which is asymptotically de Sitter and infinite in volume
in a flat slicing. In the case we are considering in this
subsection, however, the global structure of the solution is
rather different. A black hole singularity is hidden beyond $r_s$,
and the static solution does not contain any asymptotic region
that looks like (\ref{dsflat}).

\begin{figure}[tbh]
\psfrag{f}[][bl]{$\phi$}\psfrag{R}[][r]{$R$}\psfrag{0}[][bl]{$0$}
\psfrag{rs}[][r]{$r_s$} \centering \epsfysize=4cm \leavevmode
\epsfbox{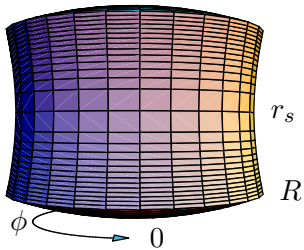} \caption{The $t=const.$ surface has the geometry
of an Einstein-Rosen bridge of the old phase which connects a pair
of bubbles of the new phase.} \label{bridge}
\end{figure}

The $t=const.$ surface has the geometry of an Einstein-Rosen
bridge of the old phase which connects a pair of bubbles of the
new phase (see Fig. \ref{bridge}). Each one of these bubbles is in
unstable equilibrium, it can either expand or contract. Let us
concentrate in one of them. If the bubble wall expands,
then its motion is  perceived as expansion from both sides of the
wall (recall that in the present case the radius $r$ decreases as
we move away from the wall in both directions). Conversely,
contraction of the bubble wall would be perceived as contraction
from both sides of the bubble wall. If one of the bubbles expands,
it eventually generates an infinitely large region of the false
vacuum phase surounding the black hole, and the metric in the false
vacuum region far away from the black hole has the asymptotic form
(\ref{dsflat}).

\begin{figure}[tbh]
\psfrag{db}[][tl]{$b$}\psfrag{s}[][r]{$s$} \psfrag{h}[][r]{$h$}
\centering \epsfysize=4cm \leavevmode \epsfbox{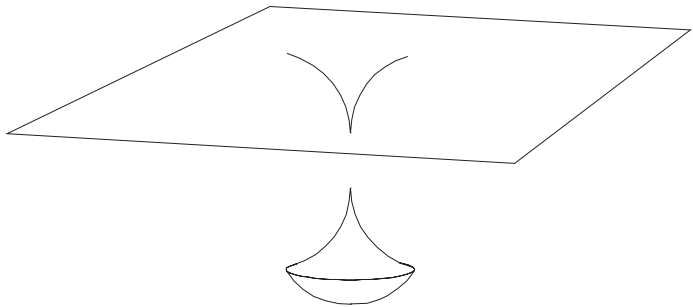}
\caption{A baby universe whith a bubble of the new phase, pinching
off a large universe filled with the old phase.} \label{gota2}
\end{figure}

This suggests the following interpretation for the static
instanton beyond the Nariai limit: it describes the thermal
production of black holes of mass $M$ [given by (\ref{themass})] in
an asymptotically de Sitter region. Initially, the throat of the black hole
connects with a compact baby universe, but this pinches off as the black hole singularity develops \cite{baby}. The baby universe
contains a bubble of the new phase in unstable equilibrium (see
Fig. \ref{gota2}). If the bubble of the new phase collapses, the
baby universe disappears into nothing. On the contrary, if the
unstable bubble expands, it ends up generating
an infinite region of the new vacuum phase, separated from an
infinite region of the old vacuum phase by a domain wall in
constant acceleration. At the center of the region of the old
vacuum phase, there is also a black hole of mass $M$.

Let us now comment on the nucleation rate. According to Eq. (\ref{B2}), this is given by
\begin{equation}
\Gamma \sim\ e^{ + S_{bh}-{A(H_o^{-1})/ 4G}}\ \sim\ e^{-\beta_o M + S_{bh}}\  e^{-{A(r_+)/ 4G}}.\quad\quad (GMH_o \ll 1)
\label{unex}
\end{equation}
Here, $S_{bh}$ is the black hole entropy. In the last step, we have used that for black holes of sufficiently low mass, the entropy of the cosmological horizon is smaller than the entropy of the original de Sitter metric by the amount $(\Delta S)_{horizon}=-\beta_o M$, as discussed in Section IV.

When we compare the previous result with Eq. (\ref{boltsu}), the last factor in Eq. (\ref{unex}) strikes us as rather unexpected. It seems to say that of all attempts at forming a black hole of mass $M$ in a region with effective de Sitter temperature $\beta_o$, only a very small fraction given by $\exp[-A(r_+)/4G]$ succeed in forming a baby universe which hosts a bubble of the new vacuum phase. Perhaps this is not unreasonable, since a baby universe which is entirely filled with the old vacuum phase would have an entropy which is higher by the amount $+A(r_+)/4G$, relative to the entropy of the baby universe containing the static bubble. This suggests that most attempts should produce a baby universe of the old phase, without a bubble of the new phase. However, we should also keep in mind that the instanton representing this alledgedly more frequent process does not exist (the solution would contain two horizons at different temperatures, and hence the Euclidean section would have a conical singularity at one of them).

Mathematically, the factor $\exp[-A(r_+)/4G]$ arises because the instanton represented in Fig. \ref{bridge} does not contain the cosmological horizon at $r_+$. The neighborhood of this horizon has been excised and replaced with the bubble of the new vacuum phase. This could mean that the interpretation given above for the instanton beyond the Nariai limit is not correct. In this interpretation, we are assuming the existence of an initial region, of size larger than the cosmological horizon, where the metric takes the approximate form (\ref{dsflat}). A cosmological horizon, and an asymptotic de Sitter region of the form (\ref{dsflat}), does develop if we let one of the unstable bubbles expand, but strictly speaking it is not present in the analytic continuation of the instanton. Clearly, the legitimacy of this interpretation deserves further investigation.

\section{Comparison with the Coleman-de Luccia action}

\begin{figure}[tbh]
\psfrag{db}[][tl]{$\hat b$}\psfrag{s}[][r]{$s$}
\psfrag{h}[][r]{$h$} \centering \epsfysize=4cm \leavevmode
\epsfbox{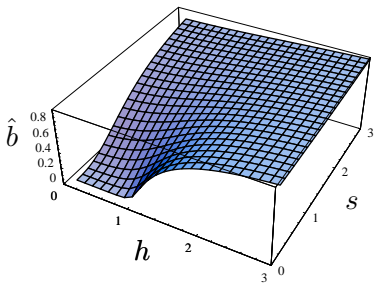} \caption{The bounce action for the Coleman-De
Luccia instantons. Here $\hat b=H_o^2 B_{\rm CDL}/8\pi^2M_p^2$,
$s=\sigma/{2M_p^2H_o}$ and $h=H_i/H_o$. The action is finite also
for upward jumps, which correspond to $h>1$.} \label{cdl}
\end{figure}

\begin{figure}[tbh]
\psfrag{db}[][tl]{$\Delta b$}\psfrag{s}[][r]{$s$}
\psfrag{h}[][r]{$h$} \centering \epsfysize=4cm \leavevmode
\epsfbox{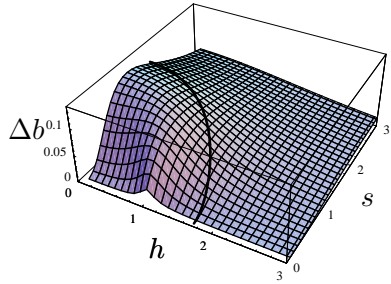} \caption{The difference between the
actions of the static and the Coleman-De Luccia instantons. Here
$\Delta b=H_o^2(B-B_{\rm CDL})/8\pi^2M_p^2$,
$s=\sigma/{2M_p^2H_o}$ and $h=H_i/H_o$. The black line on the
surface indicates the value $\sigma_{N}$ for each value of
$H_i/H_o$. Note that $\Delta b>0$ in the whole range, and
therefore the thermal activation process is always subdominant
with respect to the Coleman-de Luccia tunneling process.}
\label{deltab2}
\end{figure}

Let us now compare the action of the thermal instanton with that of the tunneling process described by
the Coleman-De Luccia (CDL) instanton. The latter is given by
(see e.g. \cite{FMSW})
\begin{equation}
B_{CDL}=12\pi^2 M_p^4\left[\frac{1}{\Lambda_o}\left(
1-b\alpha_o\right)-\frac{1}{\Lambda_i}\left( 1-b\alpha_i\right)
\right],
\end{equation}
where
\begin{equation}
\alpha_{o,i}=\frac{\epsilon}{3\sigma}\mp\frac{\sigma}{4M_p^2},
\end{equation}
and
\begin{equation}
b=\frac{1}{\sqrt{H_i^2+\alpha_i^2}}.
\end{equation}
The $\alpha$'s are related by $H_i^2+\alpha_i^2=H_o^2+\alpha_o^2$.
Using this relation and $H^2=\Lambda/3M_p^2$, $B_{CDL}$ can be
written as
\begin{equation}
B_{CDL}=\frac{8\pi^2
M_p^2}{H_o^2}\frac{1}{2\sqrt{H_i^2+\alpha_i^2}}\left[\frac{\sigma}{2M_p^2}-\frac{\epsilon}{3M_p^2}
\left( \sqrt{H_i^2+\alpha_i^2}-\alpha_i\right)\right].
\end{equation}
The values of $B$ for the static instanton and $B_{CDL}$ are
easily compared by noticing that both are of the form
$\pi/GH_o^2$, times a function of the
dimensionless parameters $s=\sigma/2M_p^2H_o$ and $h=H_i/H_o$. In
Fig. \ref{cdl} we plot the action for the CDL case, and in Fig.
\ref{deltab2} we plot the difference between the two actions. Note
that the static instanton action is larger than the CDL action in
the whole range of parameters.

\begin{figure}[tbh]
\psfrag{db}[][tl]{$r$}\psfrag{s}[][l]{$s$} \psfrag{h}[][r]{$h$}
\centering \epsfysize=6cm \leavevmode \epsfbox{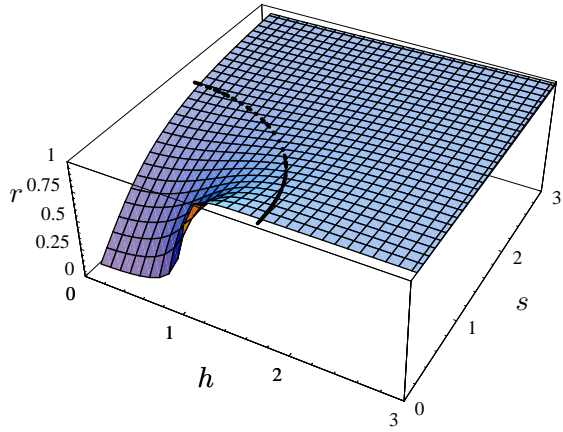}
\caption{The ratio $r=B_{CDL}/B$ between the actions of the
Coleman-De Luccia and static instantons. As in Fig. \ref{deltab2},
the black line on the surface indicates the value of
$\sigma_{N}$. The two actions are comparable for $h^2 \gtrsim 1$, and become equal only at the corner where $\sigma\to
\sigma_{N}\to 0$ and $h^2\to 3$.} \label{deltab1}
\end{figure}

As we mentioned in Section IV, jumps to a vacuum with higher energy density are also allowed \cite{leewe}. Note that for the case of upward jumps,
$h^2 \gtrsim 1$, the actions become comparable, and in fact they are equal at the corner
where $\sigma\to \sigma_{N}\to 0$ and $h^2\to 3$ (see Figs.
\ref{deltab2} and \ref{deltab1}).

\hfill\break

\section{Seeds of the new phase vs. remnants of the old phase}

\begin{figure}[tbh]
\psfrag{dbg}[][tl]{$\tilde r$}\psfrag{s}[][l]{$s$}
\psfrag{h}[][r]{$h$} \centering \epsfysize=6cm \leavevmode
\epsfbox{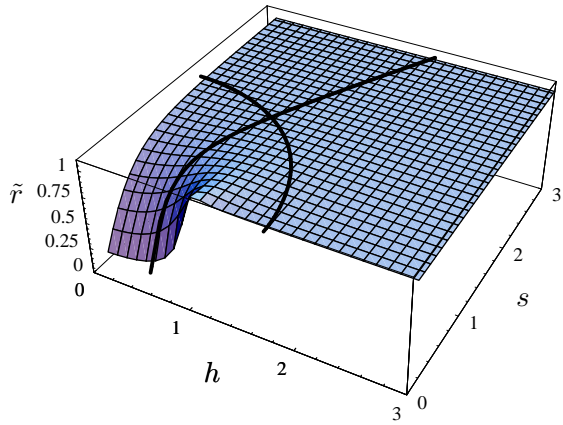} \caption{The ratio $\tilde
r=B_{CS}/B_{CR}$ between the bounce actions for the same
transition between some initial vacuum and some final vacuum,
where $h=H_{final}/H_{original}$, $B_{CS}$ corresponds to the
creation of a seed of the final vacuum and $B_{CR}$ corresponds to
the process which leaves a remnant of the original vacuum. The
right and left boundary curves correspond to the Nariai limit for
the creation of seeds and remnants respectively, $s=s_{N}$ and
$s=\tilde s_{N}$ [see Eqs. (\ref{circle}) and
(\ref{hyperbola})]. } \label{staticsobreterma}
\end{figure}

In the interpretation which we have adopted so far, the static
instanton represents the creation of pairs of critical bubbles of
the new phase embedded in the false vacuum phase \cite{game}.  We
may refer to this as the process of ``pair creation of seeds" of
the new phase. This process is analogous to pair creation of
particles (or even topological defects such as monopoles
\cite{basu}) by the expanding de Sitter background. As we showed
in Section IV, when the nucleated objects are sufficiently light,
the creation rate is simply proportional to the Boltzmann factor.

In Ref. \cite{gomberoff}, Gomberoff et al. suggested a rather
different interpretation of the same solution. The process they
considered involves a spherical bubble wall coming in from the
cosmological horizon, sweeping away the false vacuum as it moves
towards smaller radii, and replacing it with the true vacuum. The
result of this process would also be a critical bubble in unstable
equilibrium between expansion and collapse, but this time the
bubble would be a ``remnant" of the old phase rather than the seed
of the new phase. We shall thus refer to this process as ``
creation of remnants"\footnote{Gomberoff et al. used the term
``cosmological thermalon" for the process of creation of remnants.
"Thermalon" may indeed be a better word than "instanton" for
describing the static Euclidean solutions. However, this
denomination seems equally appropriate for the process of creation of
seeds, so to avoid confusion
we shall simply refer to creation of seeds or remnants.}. Even if
mathematically the Euclidean solution is the same as before, the
interpretation and background subtractions are very different. As
a consequence, the nucleation rate of such objects does {\em not}
have the same simple Boltzmann suppression form as we found in
(\ref{boltsu}) for light bubbles.

Suppose for definiteness a potential with two non-degenerate
vacua, labeled by 1 and 2, with $V(1)>V(2)>0$. The solution
representing a downward jump which leaves a remnant of vacuum 1
surrounded by vacuum 2 is the same as the instanton for an upward
jump caused by a seed of vacuum 1 which has been activated from
vacuum 2. Similarly, the upward jump which leaves a remnant of
vacuum 2 surrounded by vacuum 1 is related to downward jumps by
activation of a seed of vacuum 2 from vacuum 1. Hence
\begin{equation}
S_{CR}^{\downarrow} = S_{CS}^{\uparrow},\quad\quad
S_{CR}^{\uparrow} = S_{CS}^{\downarrow}. \label{arrows}
\end{equation}
Here, the subindex $CR$ stands for ``creation of remnants", and
$S_{CR}$ denotes the action for the cosmological thermalon
discussed by Gomberoff et al., while the subindex $CS$ refers to
``creation of seeds", and $S_{CS}$ denotes the action given in
(\ref{area}). The arrows indicate whether we are considering an
upward jump or a downward jump.

The bounce action is obtained by performing the relevant
background subtractions
\begin{equation}
B_{CR}^{\downarrow}= S_{CS}^{\uparrow}-S(1),\quad\quad
B_{CR}^{\uparrow}= S_{CS}^{\downarrow}-S(2). \label{tigger}
\end{equation}
Here $S(1)$ and $S(2)$ are the background actions, given by Eq.
(\ref{area}), with $r_+$ replaced by the corresponding de Sitter
radii $H_1^{-1}$ and $H_2^{-1}$ respectively. Before proceeding,
we should stress that since the instantons we are considering are
static and compact, then according to the discussion in Section
III the bounce actions are always given by \footnote{In fact, Ref.
\cite{gomberoff} considered a slightly different setting, where a
membrane is coupled to a three form gauge field $A_3$ and to
gravity. The term which represents the interaction of the membrane
with the gauge field takes the form $q\int A_3$, where $q$ is the
membrane charge, and the integral is over the membrane worldsheet.
It was argued in \cite{gomberoff} that $A_3$ is discontinuous
accross the membrane, and a somewhat heuristic prescription was
given to compute the contribution of $q\int A_3$ to the action and
to perform the background subtraction. The result of this
procedure, however, differs from Eq. (\ref{biarea}). Here, we
shall not try to elucidate the reason for this discrepancy. We
note, however, that the on-shell Euclidean action for the system
of a membrane coupled to $A_3$ and to gravity, and with proper
inclusion of boundary terms \cite{BT}, can be shown to be the same
as the action we have taken as our starting point
(\ref{actionstatic}) with $\rho_V$ replaced by $F^2/2$, where
$F=dA_3$ is the field strength [see e.g. Eq. (6.1) in Ref.
\cite{BT}]. Hence, the results of the present paper, which are in
principle valid for the case of vacuum decay in field theory, may
as well be valid for the case of the brane coupled to the
antisymmetric tensor field.}
\begin{equation}
B_i=-\Delta A/4G=-\Delta S \label{biarea}
\end{equation}
where $\Delta A$ is the change in the area of the horizon and
$\Delta S$ is the change in the entropy.

It is clear from (\ref{tigger}) that
\begin{eqnarray}
B_{CR}^{\uparrow}-B_{CDL}^{\uparrow} &=& B^{\downarrow}_{CS}- B_{CDL}^{\downarrow} >0, \\
B_{CR}^{\downarrow}-B_{CDL}^{\downarrow} &=& B^{\uparrow}_{CS}-
B_{CDL}^{\uparrow} >0.
\end{eqnarray}
Here, we have used the fact that
$B_{CDL}^{\uparrow}+S(2)=B_{CDL}^{\downarrow}+S(1)$, since the CDL
instanton solution is the same for upward and for downward jumps,
and all that changes is the background subtraction \cite{leewe}.
The inequalities above come from the fact that $B_{CS}$ is always
larger than $B_{CDL}$, as shown in the previous Section. It
follows that $B_{CR}$ is also always larger than $B_{CDL}$, and so
the creation of remnants is also subdominant with respect to the
tunneling process represented by the CDL instanton.

Finally, we may ask which of the two processes is more important,
the creation of seeds or the creation of remnants. From
(\ref{arrows}) we have
\begin{equation}
B_{CR}^{\uparrow}-B_{CS}^{\uparrow} =
B_{CS}^{\downarrow}-B_{CR}^{\downarrow}. \label{relat}
\end{equation}
Hence, if one of the channels is dominant for upward jumps, then
it means that the other process is dominant for downward jumps.
Fig. \ref{staticsobreterma} shows the ratio $\tilde r =
B_{CS}/B_{CR}$. Note that if $h<1$, corresponding to downward
jumps, then the process of pair creation of seeds is much more
likely than the process of pair creation of remnants. On the other
hand, for $h>1$, corresponding to upward jumps, the ratio of the
bounce actions is very close to one [although, from (\ref{relat}),
the frequency of upward jumps through creation of remnants
outweighs that of upward jumps through creation of seeds by the
same factor as downward jumps through creation of seeds outweight
downward jumps through creation of remnants.]

Finally, let us recall that the Nariai limit corresponds to
$\sigma_{N}=2 M_p^2(3H^2_{o}-H^2_{i})^{1/2}$ [see Eq.
(\ref{sigmamax})]. Here the indices $i$ and $o$ stand for the
inside and the outside of the bubble. For pair creation of seeds,
outside and inside correspond to the original vacuum and the final
vacuum respectively $H_{o}=H_{original}$ and $H_{i}=H_{final}$. In
the dimensionless variables $s=\sigma/(2 M_p^2 H_{original})$ and
$h=H_{final}/H_{original}$, the Nariai curve corresponds
to
\begin{equation}
s^2_{N}= 3- h^2. \label{circle}
\end{equation}
For pair creation of remnants with the same initial and final
states as the seeds, we have $H_{o}=H_{final}$ and
$H_{i}=H_{original}$, and the Nariai limit corresponds to
\begin{equation}
\tilde s^2_{N}= 3 h^2 - 1. \label{hyperbola}
\end{equation}
The Nariai curves $s=s_{N}$ and $s=\tilde s_{N}$, corresponding to
the circle (\ref{circle}) and the hyperbola (\ref{hyperbola}) are
also plotted in Fig. \ref{staticsobreterma}.

\section{Summary and conclusions}

De Sitter vacua are believed to be metastable at best (see e.g. \cite{susskind} for a recent discussion). It is well known that vacuum transitions from a metastable vacuum can proceed through quantum tunneling, which is described by the Coleman-de Luccia instanton. This process can take us to lower energy vacua, but also to other de Sitter vacua with a higher vacuum energy density \cite{leewe}.

Here, we have investigated an alternative process, by which critical bubbles of the new phase can be pair produced. This process is the analog of thermal activation in flat space. The mass $M$ and radius $R$ of the "seeds" of the new phase are given by Eqs. (\ref{x}) and (\ref{themass}), in terms of the initial and final vacuum energies, $\rho_o= 3M_P^2 H_o^2$ and $\rho_i=3 M_P^2 H_i^2$, and of the tension $\sigma$ of the wall separating both phases.

For $\sigma^2 < 4 M_p^4 (3H_o^2-H_i^2)$, the
geometry of the critical bubbles is the following (see Fig. 1).
Outside the bubble, the metric is Schwarzschild-de Sitter, and has
a cosmological horizon. The black hole horizon is not present,
since we are matching to an interior solution at some $R>r_s$,
where $R$ is the bubble radius and $r_s$ is the radius of the
would be black hole horizon. Inside the bubble, the metric is pure
de Sitter with curvature radius $H_i^{-1}$. For $GMH_o \ll 1$, the
nucleation rate is proportional to the Boltzmann
factor
\begin{equation}
\Gamma \sim e^{-\beta_o M},
\label{conc}
\end{equation}
as would be expected from simple thermodinamical arguments. Here $\beta_o=2\pi/H_o$ is the inverse de Sitter temperature of the old vacuum phase.

For $\sigma^2=4 M_p^4(3H_o^2-H_i^2)$ the metric outside of the
bubble corresponds to the Nariai limit of the Schwarzschild-de
Sitter solution, for which the black hole and cosmological
horizons have the same size.
Beyond the Nariai limit, i.e. for $\sigma^2>4
M_p^4(3H_o^2-H_i^2)$, the asymptotic form of the solution changes
quite drastically \cite{gomberoff}. The static solution with a
pair of critical bubbles has a black hole horizon instead of a
cosmological horizon. The interpretation of such solution is less
clear than in the case $\sigma^2<4 M_p^4(3H_o^2-H_i^2)$, but we
have argued that it may correspond to the creation of a baby
universe containing a bubble of the new phase. The nucleation rate
is formally given by (\ref{unex}), and does not have the simple
form (\ref{conc}) even in the case when the mass $M$ is small
(here, $M$ is the mass of the black hole connecting the asymptotic
region of the old phase with the baby universe).

We have compared the process of thermal activation of seeds to an alternative process recently suggested by Gomberoff et al. \cite{gomberoff}, by which most of space would suddenly jump to the new vacuum phase, leaving only a pair of critical bubbles as remnants of the old phase. These could subsequently collapse into black holes, with the net result that the vacuum ``dark" energy, is transformed into cold ``dark matter" in the form of black holes.
We find that for downward jumps, this process is subdominant with respect to the thermal activation of seeds of the new vacuum. For upward jumps, the bounce actions are comparable, and in fact the creation of remnants may be slightly favoured with respect to the creation of seeds (although when we are going up in energy we are not transforming dark energy into dark matter, but simply increasing both of them!).

Also, we have compared the rate of nucleation of critical bubbles by thermal activation with the rate of bubble nucleation by quantum tunneling, described by the Coleman-de Luccia (CDL) instanton. The CDL instanton always has a lower bounce action than the process of thermal activation of seeds or remnants. Thus, even if thermal activation is possible, it appears that jumps between neighboring vacua will be more frequent through quantum tunneling. For the case of upward jumps, however, the corresponding actions are comparable (see Fig. 6).
Since the action for thermal activation is higher than that for tunneling, one should ask whether there are any situations where the former process may nevertheless be relevant. Note that if the bubble wall carries some internal degrees of freedom, their entropy will be accounted for in the prefactor which accompanies the leading expression $e^{-B}$ for the nucleation rate. It is clear from Eq. (\ref{hite}) that the temperature the bubble wall can be very high if the wall tension and the vacuum energies in the two phases are suitably adjusted. Hence the entropy of the internal degrees of freedom can be very high, making up perhaps for the difference in actions. Investigation of this possibility is left for further research.
\footnote{In Ref. \cite{game}, we speculated that thermal activation may be relevant for the process of multiple brane nucleation. As noted in \cite{FMSW}, coincident branes carry a number of degrees of freedom which grows nonlinearly with the number of branes. Because of that, the entropy factors due to the fields living on the branes may greatly enhance the nucleation rates, and it may be more probable to nucleate a bubble bounded by a whole stack of branes, than a bubble bounded by a single brane. Note, however, that in four dimensions (and after the dilaton is stabilized) the interactions amongst 2-branes are repulsive. Because of that, the CDL instanton for multiple brane nucleation may not exist, since the stack of branes does not hold together. On the other hand, at sufficiently high temperature, the branes may attract each other because of thermal symmetry restoration. Hence, it is conceivable that the CDL instanton may not exist while the thermal instanton does \cite{game}. This possibility seems rather exotic, but it may well be realized in certain regions of parameter space.}

\section*{Acknowledgements}

J.G. is grateful to the organizers of the 8th Peyresq meeting for
hospitality, and to R. Emparan and M. Porrati for useful
discussions. A.M. was supported by a postdoctoral fellowship of
the Ministerio de Educaci\'on y Cultura, Spain. The work by J.G.
was supported by CICYT Research Projects FPA2002-3598,
FPA2002-00748, and DURSI 2001-SGR-0061.

\end{document}